\documentclass[twocolumn,preprintnumbers,amsmath,amssymb,aps]{revtex4}
\usepackage{graphicx}
\usepackage{dcolumn}
\usepackage{bm}
\usepackage{float}
\usepackage{color}
\usepackage{soul}
\usepackage{mathtools}

\begin{document}

\title{Quantum coherence assisted dynamical phase transition}

\author{Bao-Ming Xu}
\email{xbmv@bit.edu.cn}

\affiliation{Shandong Key Laboratory of Biophysics, Institute of Biophysics, Dezhou University, Dezhou 253023, China}%

\date{Submitted \today}

\begin{abstract}
Quantum coherence will undoubtedly play a fundamental role in understanding the dynamics of quantum many-body systems, thereby to reveal its genuine contribution is of great importance. In this paper, we specialize our discussions on the one-dimensional transverse field quantum Ising model initialized in the coherent Gibbs state, and investigate the effects of quantum coherence on dynamical phase transition (DQPT). After quenching the strength of the transverse field, the effects of quantum coherence are studied by Fisher zeros, rate function and winding number. We find that quantum coherence not only recovers the traditional DQPT related to quantum phase transition, but also generates some entirely new DQPTs which are independent of equilibrium quantum critical point. In these entirely new QDPTs, the line of Fisher zeros cuts the imaginary axis twice, i.e., there are two critical modes, one makes the winding number jump down but another makes it jump up. We also find that the rate function can not be used to describe DQPT at high temperature, because the critical mode no longer dominates. This work sheds new light on the fundamental connection between quantum critical phenomena and quantum coherence.
\end{abstract}

\maketitle

\section{Introduction}

Experiments with ultracold atom \cite{Levin2012,Yukalov2011,Bloch2008,Greiner2002} and ion-trap \cite{Porras2004,Kim2009,Jurcevic2014} provide opportunities to manipulate and control the system, both in space and in time, with an unprecedented accuracy as compared to any solid-state counterpart, preserving coherence over long time scales. With these experimental advances, an in-depth study of the dynamics of quantum many-body systems far from thermodynamic equilibrium becomes possible. Specifically, realizing the quantum time evolution beyond the thermodynamic equilibrium description as well as the dynamics of out-of-equilibrium quantum many-body criticality, has recently attracted considerable attention \cite{Eisert2015}. Dynamical quantum phase transitions (DQPTs) have recently emerged as an interesting phenomenon within this regime \cite{Marino2022,Heyl2019,Heyl2018,Zvyagin2016}, which opens the possibility of studying fundamental questions such as scaling and universality in quantum real-time evolution.

DQPTs occur in the dynamics of a closed quantum many-body systems after quenching a set of control parameters of its Hamiltonian, and generally refer to two largely independent concepts. The first one, DQPT-I \cite{Eckstein2008,Eckstein2009,Moeckel2008,Moeckel2010,Sciolla2010,Sciolla2011,Gambassi2013,
Sciolla2013,Maraga2015,Chandran2013,Smacchia2015,Halimeh2017,Mori2018,Zhang2017,Muniz2020,Smale2019,Tian2020}, resembls the equilibrium Landau theory: A system undergoes a dynamical phase transition at the critical value of the control parameter if the long-time limit of physically relevant observables, namely the order parameter, is finite for one phase, whereas it vanishes for another. These asymptotic values are somehow connected with prethermalization \cite{Gring2012,Langen2015,Neyenhuis2017}. Furthermore, DQPT-I also entails criticality in the transient dynamics of the order parameter and two-point correlators before reaching the steady state, giving rise to effects such as dynamic scaling and aging, which have been investigated theoretically \cite{Chiocchetta2015,Marcuzzi2016,Chiocchetta2017} and also observed experimentally \cite{Nicklas2015}.
The second concept, DQPT-II \cite{Heyl2013,Heyl2014,Andraschko2014,Vajna2014,Heyl2015,Budich2016,Bhattacharya2017,Heyl2017,Cao2020,Cao2023,Cheraghi2023}, generalizes the nonanalytic behavior of the free energy at a phase transition in the thermodynamic limit to the out-of-equilibrium case. To this end, the Loschmidt echo has been introduced as a dynamical analog of a free energy per particle. This is therefore an inherently dynamical effect, taking place before the system has reached equilibration at all. There is no order parameter in the conventional sense linked to this kind of DQPT, and the creation of different phases is not immediately apparent from the nature of the effect. In the seminal paper \cite{Heyl2013}, DQPT-II were originally proposed using the nearest-neighbor transverse-field Ising model, however it was soon shown to hold for a wide variety of models, ranging from topological matter \cite{Vajna2015,Schmitt2015,Jafari2016,Jafari2017,Sedlmayr2018,Jafari2018,Zache2019,Maslowski2020,Okugawa2021,Cao2022}, non-Hermitian systems \cite{Zhou2018,Zhou2021a,Mondal2022}, Floquet systems \cite{Sharma2014,Yang2019,Zamani2020,Zhou2021b,Jafari2021,Hamazaki2021,Zamani2022,Luan2022,Jafari2022} to time crystals \cite{Kosior2018a,Kosior2018b}. Moreover, DQPT-II were experimentally observed in trapped ion quantum simulators \cite{Jurcevic2017}, superconducting qubits \cite{Guo2019}, and other platforms \cite{Flaschner2018,Tian2019,Wang2019,Xu2020}. Both kinds of DQPTs may appear in the same models, for example the long-range or fully connected transverse-field Ising model \cite{Zunkovic2018,Lang2018a,Halimeh2017,Stauber2017}, or the Rabi model \cite{Puebla2020}, although the strict connection between them is still an open question \cite{Lang2018b,Weidinger2017,Hashizume2022,Sehrawat2021,Lerose2019,Zunkovic2016,Corps2022,Corps2023a,Corps2023b}. Inspired by the physics of fully-connected systems and the phases demarcated by an excited-state quantum phase transition \cite{Cejnar2021,Cejnar2006,Caprio2008,Stransky2014,Stransky2015,Stransky2019}, DQPTs-I and DQPTs-II were argued to be triggered by the behavior of an operator which is constant only in some the excited-state quantum phases \cite{Corps2022,Corps2023a,Corps2023b}. In this paper, we will be concerned with DQPTs-II only, which we will henceforth simply denote DQPTs.

On the other hand, quantum coherence, as a fundamental concept in quantum mechanics that sets it apart from classical physics, will undoubtedly play an important and pivotal role in understanding the dynamics of quantum many-body systems. Quantum coherence in many-body systems embodies the essence of entanglement and is an essential ingredient for a plethora of physical phenomena. It has been found that many-body coherence can eliminate the well-known Coulomb staircase and lead to strong negative differential resistance, thereby controlling electron transport \cite{Hang2023}. If transport is absent, the system cannot act as a heat bath for its constituent parts, resulting in a many-body localized phase \cite{Schreiber2015}, and that is why quantum coherence can be used as an order parameter to detect the well-studied ergodic to many-body localized phase transition \cite{Dhara2020,Styliaris2019}. Furthermore, a rigorous connection between quantum coherence and quantum chaos was also established \cite{Anand2021}. It is well known that a phase transition is a fundamental change in the state of a system when one of the parameters of the system passes through its critical point characterized by diverging correlation length and relaxation time (vanishing energy gap). This critical slowing down implies that it is impossible to pass the critical point at a finite speed without exciting the system. In other words, quantum coherence is inevitable in quantum phase transitions, whose dynamics obeys Kibble-Zurek mechanism\cite{Dziarmaga2022} or power law \cite{Brauna2015}. Because of this, quantum coherence can be employed to detect and characterize quantum phase transitions \cite{Cakmak2015,Karpat2014,Girolami2014,Malvezzi2016,Chen2016,Li2016,Li2020,Mao2021,Swan2020,Lv2022,Pires2021,Wanf2022}. Moreover, a contribution of quantum coherence to quantum critical phenomena has also been found in the framework of quantum thermodynamics \cite{Xu2018,Santini2023,Xu2023}. There is no doubt that quantum coherence also plays a fundamental role on DQPTs, but it has not been studied until now, to the best of our knowledge.

In this paper, we aim at revealing some unique effects of quantum coherence on DQPTs with the aid of large deviation theory \cite{Touchette2009,Gambassi2012}. To accomplish this, we specialize our discussions to the quenched one-dimensional transverse field quantum Ising model, one of two prototypical models to understand the quantum phase transition \cite{Sachdev2011}. The system is first prepared in the coherent Gibbs state \cite{Xu2018,Lostaglio2015,Kwon2018}, an extreme example of coherent resources in quantum thermodynamics, and then quenched by suddenly changing the transverse field, a common way to drive an isolated quantum system out of equilibrium. After quenching, Fisher zeros \cite{Heyl2013,Fisher1965}, the return probability of the time evolved wavefunction to its initial state (i.e., Loschmidt echo) and the winding number are derived. With these concepts at hand, we then discuss the effects of quantum coherence on DQPTs.

This paper is organized as follows: In the next section, we introduce the quenched one-dimensional transverse Ising model and briefly review the crucial method for the analytical solution, which will be used in the following. In Sec. III, Loschmidt echo, Fisher zeros and winding number are derived. DQPT for the thermal equilibrium state are discussed in Sec. IV. The effects of quantum coherence on DQPT are discussed in Sec. V. Finally, Sec. VI closes the paper with some concluding remarks.

\section{Quenched one-dimensional Transverse Field Ising model}
The Hamiltonian of the one-dimensional transverse field Ising model is
\begin{equation}\label{}
  \hat{H}\bigl(\lambda\bigr)=-\frac{J}{2}\sum_{j=1}^{N}\bigl[\hat{\sigma}_j^z\hat{\sigma}_{j+1}^z+\lambda\hat{\sigma}_j^x\bigr],
\end{equation}
where $J$ is longitudinal spin-spin coupling, $\lambda$ is a dimensionless parameter measuring the strength of the transverse field with respect to the longitudinal spin-spin coupling. In this work, we set $J=1$ as the overall energy scale and only consider $\lambda\geq0$ without loss of generality. $\hat{\sigma}^{\alpha}_{j}$ $(\alpha=x,y,z)$ is the spin-1/2 Pauli operator at lattice site $j$ and the periodic boundary conditions are imposed as $\hat{\sigma}^{\alpha}_{N+1}=\hat{\sigma}^{\alpha}_{1}$. Here we only consider that $N$ is even. The one-dimensional quantum Ising model is the prototypical, exactly solvable example of a quantum phase transition, with a quantum critical point at $\lambda_c=1$ separating a quantum paramagnetic phase at $\lambda>\lambda_c$ from a ferromagnetic one at $\lambda<\lambda_c$.

After Jordan-Wigner transformation and Fourier transforming, the Hamiltonian becomes a sum of two-level systems \cite{Russomanno2012}:
\begin{equation}\label{}
\hat{H}(\lambda)=\sum_k\hat{H}_k(\lambda).
\end{equation}
Each $\hat{H}_k(\lambda)$ acts on a two-dimensional Hilbert space generated by $\{\hat{c}^\dag_k\hat{c}^\dag_{-k}|0\rangle,~|0\rangle\}$, where $|0\rangle$ is the vacuum of the Jordan-Wigner fermions $\hat{c}_k$, and can be represented in that basis by a $2\times2$ matrix
\begin{equation}\label{}
 \hat{H}_k(\lambda)=(\lambda-\cos k)\hat{\sigma}^z+\sin k\hat{\sigma}^y,
\end{equation}
where $k=(2n-1)\pi/N$ with $n=1\cdots N/2$, corresponding to antiperiodic boundary conditions for $N$ is even. The instantaneous eigenvalues are $\varepsilon_{k}(\lambda)$ and $-\varepsilon_{k}(\lambda)$ with
\begin{equation}\label{}
\varepsilon_{k}(\lambda)=\sqrt{(\lambda-\cos k)^2+\sin^2k}.
\end{equation}
The corresponding eigenvectors are
\begin{equation}\label{}
|\varepsilon^+_{k}(\lambda)\rangle=
\bigl[i\sin\theta_k(\lambda)+\cos\theta_k(\lambda)\hat{c}^\dag_k\hat{c}^\dag_{-k}\bigr]|0\rangle
\end{equation}
and
\begin{equation}\label{}
|\varepsilon^-_{k}(\lambda)\rangle=\bigl[\cos\theta_k(\lambda)+i\sin\theta_k(\lambda)\hat{c}^\dag_k\hat{c}^\dag_{-k}\bigr]|0\rangle,
\end{equation}
respectively, where $\theta_k(\lambda)$ is determined by the relation
\begin{equation}\label{}
e^{i\theta_k(\lambda)}=\frac{\lambda-\varepsilon_{k}(\lambda)-e^{-ik}}{\sqrt{\sin^2k+[\lambda-\cos k-\varepsilon_{k}(\lambda)]^2}}.
\end{equation}

In order to explore the effects of quantum coherence, we consider the system to be initially in the coherent Gibbs state
\begin{equation}\label{}
  |\psi(0)\rangle=\bigotimes_{k}|\psi(0)\rangle_k
\end{equation}
with
\begin{equation}\label{}
  |\psi_k(0)\rangle=\sqrt{\frac{e^{-\beta\hat{H}_k(\lambda)}}{Z_{k}(\lambda)}}
  \biggl(|\varepsilon^+_{k}(\lambda)\rangle+e^{i\phi_k}|\varepsilon^-_{k}(\lambda)\rangle\biggr),
\end{equation}
where $\beta=1/T$ is the inverse of the temperature (we have set Boltzmann constant $k_B=1$), $Z_k(\lambda)=2\cosh[\beta\varepsilon_{k}(\lambda)]$ and $\phi_k$ refer to the partition function and the relative phase for mode $k$. In this paper, we consider that all the modes have the same relative phase for simplicity, i.e., $\phi_k=\phi$ for all $k$. It is worth noting that even though all the modes have the same relative phase, the different energy levels have different relative phases. Experimentally, spin Hamiltonians can be implemented using ultracold atoms in optical lattices with excellent isolation from the environment \cite{Schauss2018}. The challenge in realizing the coherent Gibbs state is to collectively and synchronously manipulate all the lattices, not only to achieve the superposition of different energy levels but also to regulate their phases. Coherent Gibbs state is energy indistinguishable with the corresponding Gibbs state or thermal equilibrium state $\hat{\rho}_{th}(0)=\bigotimes_{k}\frac{e^{-\beta\hat{H}_k(\lambda)}}{Z_{k}(\lambda)}$ because they have the same diagonal elements, and in this sense we call parameter $\beta$ in $|\psi(0)\rangle$ the ``effective temperature". Coherent Gibbs state is usually used to investigate the role of quantum coherence in thermodynamics because it has a direct correspondence with Gibbs state. This direct correspondence allows us to separate quantum properties from thermodynamics and study the tradeoffs between quantum and thermal fluctuations. Because of this favorable properties, coherent Gibbs state is used to study the effects of quantum coherence on many-body systems and distinguish the contributions of quantum and thermal fluctuations on DQPT in this paper. It is necessary to emphasize that this paper only focuses on the quantum coherence between energy levels because quantum coherence depends on the choice of eigenbasis. At zero temperature $\beta\rightarrow\infty$, coherent Gibbs state is the ground state, with on quantum coherence. At the other extreme, i.e., infinite temperature $\beta\rightarrow0$, the Gibbs state is a equal-probability-amplitude superposition of all the energy levels. Finally, we emphasize that although coherent Gibbs states is a pure state, the results obtained from it can also be demonstrated by the mixed state of $\rho(0)=(1-p)\hat{\rho}_{th}(0)+p|\psi(0)\rangle\langle\psi(0)|$ with $p\in[0,1]$ being the probability of coherent Gibbs state.

A common way to take a many-body system out of equilibrium is by an abrupt change of a local or global parameter of the Hamiltonian, this is commonly referred to as a ``sudden quench". At time $t=0$, the transverse field is suddenly changed from $\lambda$ to $\lambda'$. We assume that this process is so sudden that the system state has no time to change. The pre- and post-quench Hamiltonians are not commutative, i.e., $[\hat{H}(\lambda),\hat{H}(\lambda')]\neq0$. It is well known that this non-commutation leads to quantum fluctuations that govern the properties of the ground state of the system, such as spin-polarization or magnetic susceptibility, causing a order-disorder symmetry-breaking transition at the critical point. Beyond this equilibrium paradigm, a novel dynamical behavior referred to the dynamical quantum phase transition (DQPT) can also arise from these quantum fluctuations, which is the main concern of this paper. After quenching, the dynamics of the system is governed by the post-quench Hamiltonian $\hat{H}(\lambda')$, and the system state at time $t$ is determined by ($\hbar=1$)
\begin{equation}\label{}
  |\psi(t)\rangle=\hat{U}(t)|\psi(0)\rangle=e^{-i\hat{H}(\lambda')t}|\psi(0)\rangle.
\end{equation}
It is worth noting that if we do not consider quench, i.e., $\hat{H}(\lambda')=\hat{H}(\lambda)$, the system state will also evolve which is determined by $|\psi(t)\rangle=e^{-i\hat{H}(\lambda)t}|\psi(0)\rangle$, but DQPT can not occur. This result confirms that DQPT is caused by quantum fluctuations that arise from the non-commutation between $\hat{H}(\lambda)$ and $\hat{H}(\lambda')$. However, it must be emphasized that some DQPTs do not depend on equilibrium quantum phase transitions, which will be discussed in the following.

\begin{figure*}
\begin{center}
\includegraphics[width=16cm]{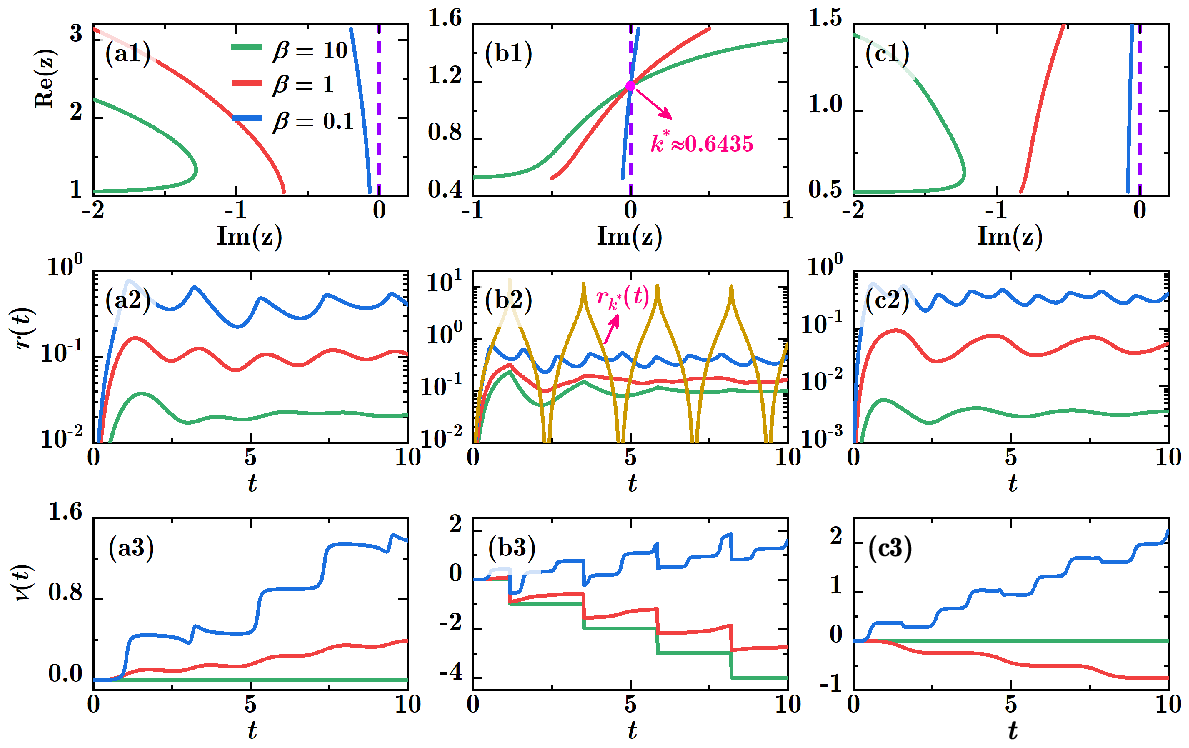}
\caption{ (Color online) Lines of Fisher zeros (a1-c1), the time evolution of rate function $r(t)$ (a2-c2) and the winding number $\nu(t)$ (a3-c3) for a quench within the ferromagnetic phase $\lambda=0\rightarrow\lambda'=0.5$ (a1-a3) and paramagnetic phase $\lambda=1.5\rightarrow\lambda'=2$ (c1-c3), and across the equilibrium quantum critical point $\lambda=0.5\rightarrow\lambda'=2$ (b1-b3) at temperature $\beta=10$ (green), $\beta=1$ (red) and $\beta=0.1$ (blue). The purple dashed lines in (a1-c1) are the imaginary axis of the Fisher zeros. The corresponding critical mode is $k^*\approx0.6435$. The yellow curves in (b2) is the time evolution of the critical rate function $r_{k^*}(t)$. For all the cases, $\phi=0$ and $n=0$.}
\label{figure1}
\end{center}
\end{figure*}

\section{Loschmidt echo and Fisher zeros}
The central object within the theory of DQPTs is the Loschmidt amplitude
\begin{equation}\label{overlap}
  \mathcal{G}(t)=\langle\psi(0)|\psi(t)\rangle=\prod_k\langle\psi_k(0)|e^{-i\hat{H}_k(\lambda')t}|\psi_k(0)\rangle,
\end{equation}
which quantifies the deviation of the time-evolved state from the initial condition.
It should be noted that when $\phi=0$, $\langle\psi(0)|\psi(t)\rangle=\mathrm{Tr}[\hat{\rho}_{th}(0)\hat{U}(t)]$ because $\langle\varepsilon^+_{k}(\lambda)|e^{-i\hat{H}_k(\lambda')t}|\varepsilon^-_{k}(\lambda)\rangle
+\langle\varepsilon^-_{k}(\lambda)|e^{-i\hat{H}_k(\lambda')t}|\varepsilon^+_{k}(\lambda)\rangle=0$, which recovers the generalized Loschmidt overlap amplitude in Refs. \cite{Bhattacharya2017,Heyl2017}. In other words, the results of thermal equilibrium state can be simulated by coherent Gibbs state with $\phi=0$. In the following, we will adjust $\phi$ to realize the comparison with thermal equilibrium state. The return probability $\mathcal{L}(t)$ associated with the amplitude $\mathcal{G}(t)$:
\begin{equation}\label{}
 \mathcal{L}(t)=|\mathcal{G}(t)|^2
\end{equation}
will be termed the Loschmidt echo. Loschmidt echo or the return probability obeys a large deviation form $\mathcal{L}(t)\sim e^{-Nr(t)}$ with a rate function
\begin{equation}\label{}
r(t)=-\lim_{N\rightarrow\infty}\frac{1}{N}\ln\mathcal{L}(t),
\end{equation}
which can be understood as a dynamical analog of the thermal free energy. In the thermodynamic limit one can derive an exact result for $r(t)$: \begin{widetext}
\begin{equation}\label{}
  r(t)=-\frac{1}{\pi}\mathrm{Re}\Biggl[\int_0^\pi dk \ln\biggl[\cos[\varepsilon_{k}(\lambda')t]
  +i\sin[\varepsilon_{k}(\lambda')t]\Bigl[\cos(2\Delta_{\theta_k})\tanh[\beta\varepsilon_{k}(\lambda)]
  +\frac{2}{Z_k(\lambda)}\sin\phi\sin\bigl(2\Delta_{\theta_k}\bigr)\Bigr]\biggr]\Biggr],
\end{equation}
\end{widetext}
where $\Delta_{\theta_k}=\theta_k(\lambda)-\theta_k(\lambda')$.
A DQPT will occur when $r(t)$ exhibits a nonanalyticity at some times $t^*$, namely critical times.

Besides the rate function $r(t)$, we will also investigate the real-valued, noncyclic geometric phase of the Loschmidt overlap amplitude $\mathcal{G}_k(t)$, which is defined as
\begin{equation}\label{}
  \phi_k^G(t)=\phi_k(t)-\phi_k^D(t)
\end{equation}
with
\begin{equation}\label{}
  \phi_k(t)=\arg[\langle\psi_k(0)|e^{-i\hat{H}_k(\lambda')t}|\psi_k(0)\rangle]
\end{equation}
being the total phase of $\mathcal{G}_k(t)$ and
\begin{equation}\label{}
  \phi_k^D(t)=-\int_0^t dt'\langle\psi_k(0)|\hat{H}_k(\lambda')|\psi_k(0)\rangle.
\end{equation}
being the dynamical phase. At a given time, the winding number of the geometric phase in the first Brillouin zone can be further employed to construct a dynamical topological order parameter (DTOP), which is defined as
\begin{equation}\label{}
  \nu(t)=\frac{1}{2\pi}\int_0^\pi\frac{\partial\phi_k^D(t)}{\partial k}dk.
\end{equation}
It takes a quantized jump whenever the evolution of the system passes through a critical time of the DQPT.

A powerful method to analyse the nonanalyticity of $r(t)$ is Fisher zeros \cite{Fisher1965}. In order to apply this concept, we should expend time $t$ into the complex plane and focus on the boundary partition function
\begin{equation}\label{Zz}
  Z(z)=\langle\psi(0)|e^{-z\hat{H}(\lambda')}|\psi(0)\rangle=\prod_k\langle\psi_k(0)|e^{-z\hat{H}_k(\lambda')}|\psi_k(0)\rangle
\end{equation}
where $z\in\mathbb{C}$. For imaginary $z=it$ this just describes the overlap amplitude of Eq. (\ref{overlap}). For real $z=R$ it can be interpreted as the partition function of the field theory described by $\hat{H}(\lambda')$ with boundaries described by boundary states $|\psi(0)\rangle$ separated by $R$ \cite{LeClair1995}. The Fisher zeros are the values of $z$ that make $Z(z)=\langle\psi(0)|e^{-z\hat{H}(\lambda')}|\psi(0)\rangle=0$. Because of Eq. (\ref{Zz}), a zero in $Z(z)$ is equivalent to finding at least one critical momentum $k$ and one critical $z$ where $\langle\psi_k(0)|e^{-z\hat{H}_k(\lambda')}|\psi_k(0)\rangle = 0$. In the thermodynamic limit the zeros of the partition function in the complex plane coalesce to a family of lines labeled by a number $n\in\mathbb{Z}$
\begin{widetext}
\begin{equation}\label{}
z_n(k)=\frac{1}{2\varepsilon_{k}(\lambda')}\Biggl[\ln\bigg\vert
  \frac{e^{-\beta\varepsilon_{k}(\lambda)}\cos^2\Delta_{\theta_k}+e^{\beta\varepsilon_{k}(\lambda)}\sin^2\Delta_{\theta_k}-\sin\phi\sin(2\Delta_{\theta_k})}
  {e^{-\beta\varepsilon_{k}(\lambda)}\sin^2\Delta_{\theta_k}+e^{\beta\varepsilon_{k}(\lambda)}\cos^2\Delta_{\theta_k}+\sin\phi\sin(2\Delta_{\theta_k})}
  \bigg\vert+i(2n+1)\pi\Biggr].
\end{equation}
\end{widetext}
Then, there is a series of critical times $t^*_n$
\begin{equation}\label{critical time}
  t^*_n=\frac{1}{2\varepsilon_{k^\star}(\lambda')}(2n+1)\pi
\end{equation}
depending on the corresponding critical mode $k^*$ determined by equation
 \begin{equation}\label{critical mode}
  -\tanh[\beta\varepsilon_{k}(\lambda)]\cot(2\Delta_{\theta_k})=\sin\phi.
\end{equation}
It can be seen that the critical mode is determined by the interplay of temperature, relative phase and quench. For the thermal equilibrium state, i.e., $\phi=0$, the critical mode $k^*$ is determined by equation $\cos(2\Delta_{\theta_k})=0$, independent of temperature, and thereby rendering the critical time also independent of temperature \cite{Bhattacharya2017,Heyl2017}.

\begin{figure*}
\begin{center}
\includegraphics[width=16cm]{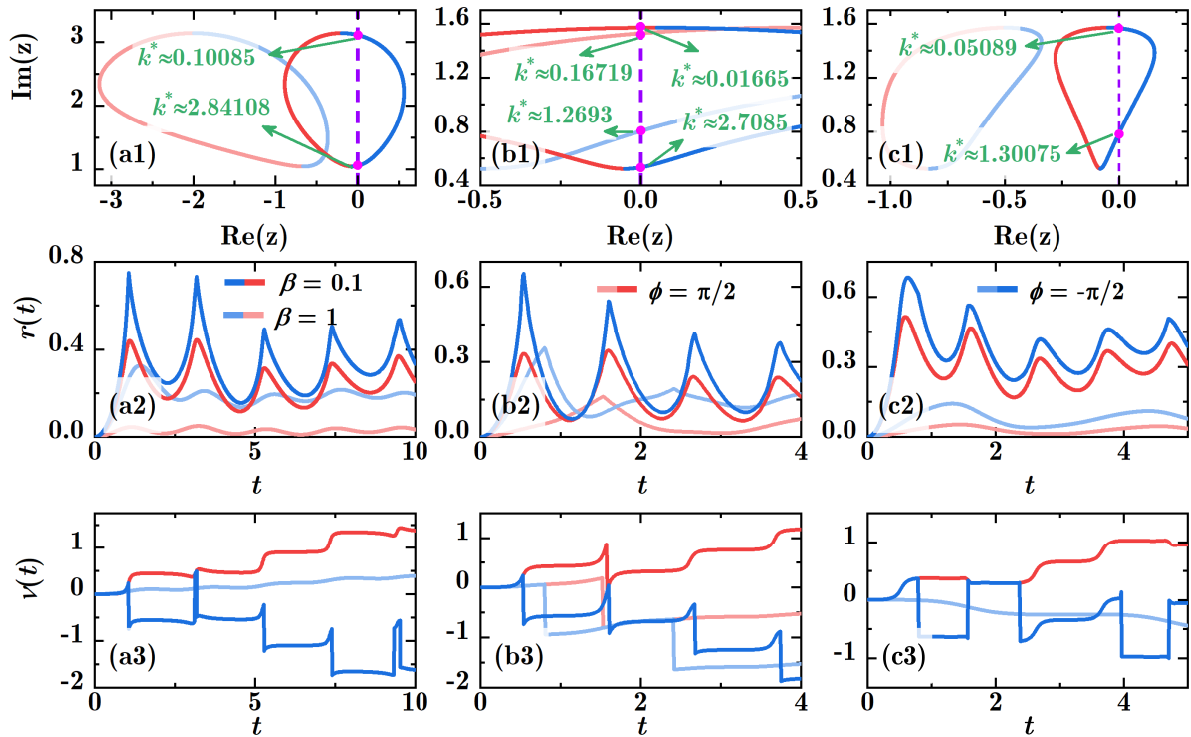}
\caption{(Color online) Lines of Fisher zeros (a1-c1), the time evolution of rate function $r(t)$ (a2-c2) and the winding number $\nu(t)$ (a3-c3) for a quench within the ferromagnetic phase $\lambda=0\rightarrow\lambda'=0.5$ (a1-a3) and paramagnetic phase $\lambda=1.5\rightarrow\lambda'=2$ (c1-c3), and across the equilibrium quantum critical point $\lambda=0.5\rightarrow\lambda'=2$ (b1-b3) at temperatures $\beta=1$ (light red and light blue curves) and $\beta=0.1$ (dark red and dark blue curves). The relative phases are $\phi=\pi/2$ (red) and $-\pi/2$ (blue). The purple lines in (a1-c1) are the imaginary axis of the Fisher zeros. The critical modes is given (see the green arrows and green values). It should be noted that the winding numbers for $\phi=\pi/2$ and $-\pi/2$ are coincide with each other at media temperature $T=1$ (the light blue and light red curves in (a3) and (c3)). For all the cases, $n=0$.}
\label{figure2}
\end{center}
\end{figure*}

\section{DQPT for the thermal equilibrium state}
In this section, we investigate DQPTs for thermal equilibrium state letting $\phi=0$. At low temperatures, the system mainly populates the ground state. If the quench crosses the quantum critical point at which the equilibrium quantum phase transition occurs, the line of Fisher zeros will cut $\mathrm{Im}(z)$ axis [see the green curve in Fig. \ref{figure1}(b1)], giving rise to nonanalytic behavior (cusp singularity) of $r(t)$ which implies DQPT occurring [see the green curve in Fig. \ref{figure1}(b2)] where the winding number $\nu(t)$ discontinuously jumps down at critical times as expected [see the green curve in Fig. \ref{figure1}(b3)]. As the temperature rises, the line of Fisher zeros still cuts $\mathrm{Im}(z)$ axis [see the blue and red curves in Fig. \ref{figure1}(b1)]. In fact, critical time and critical mode do not depend on the temperature \cite{Bhattacharya2017,Heyl2017} [see the purple dot in Fig. \ref{figure1}(b1)]. As the results, DQPT can still be observed at non-zero temperatures [see the red and blue curves in Fig. \ref{figure1}(b3)]. But we find that the cusp singularity of $r(t)$ can only be observed at temperatures $T\lesssim1$ and will be destroyed at high temperature $T\sim10$ [see the red and blue curves in Fig. \ref{figure1}(b2)]. To understand why the cusp singularity of $r(t)$ disappear at high temperature, we compare the rate function of the critical mode $k^{*}$, namely critical rate function $r_{k^*}(t)=\ln|\langle\psi_{k^*}(0)|e^{-i\hat{H}_{k^*}(\lambda')t}|\psi_{k^*}(0)\rangle|^2$ (the yellow curve), with rate functions $r(t)$ in Fig. \ref{figure1}(b2). At low temperature, rate function $r(t)$ (although it is related to all modes) resonates with $r_{k^*}(t)$, or the return probability $|\langle\psi(0)|e^{-i\hat{H}(\lambda')t}|\psi(0)\rangle|^2$ resonates with $|\langle\psi_{k^*}(0)|e^{-i\hat{H}_{k^*}(\lambda')t}|\psi_{k^*}(0)\rangle|^2$. In other words, it is the critical mode $k^*$ that mainly determines the short-time dynamics of the system, thereby displaying its unique singularity. As temperature rises, other modes come into play, weakening the dominance of the critical model [see the red curve in Fig. \ref{figure1}(b2)]. At sufficiently high temperatures, all modes are at the same level, and the dominance of the critical mode is completely deprived [see the blue curve in Fig. \ref{figure1}(b2)], therefore no cusp singularity of $r(t)$ can be observed. If a quench is performed within the same phase (ferromagnetic or paramagnetic), the line of Fisher zeros does not intersect with $\mathrm{Im}(z)$ axis [see Fig. \ref{figure1}(a1) and Fig. \ref{figure1}(c1)], and no DQPT can be observed [see Fig. \ref{figure1}(a2-a3) and Fig. \ref{figure1}(c2-c3)].

According to the discussion above, we have known that DQPT can survive at medial temperature from the Loschmidt echo perspective, which is expected in Refs. \cite{Bhattacharya2017,Heyl2017}. However, DQPT, even at low temperatures, can not be detected by the rate function of work distribution \cite{Abeling2016}. The rate function of work distribution, just as shown in the seminal paper \cite{Heyl2013}, can be used to detect DQPT only at absolute zero temperature where the system is completely stay at the ground state. From work distribution point of view, DQPT is so fragile that even a little thermal fluctuation is enough to disrupt it, but from the Loschmidt echo perspective, it is stable and able to resist thermal fluctuations to some extent. This seems contradictory, but it is not the case. In the work distribution perspective, it is generally the probability of null-work being used to detect DQPT. For the initial equilibrium state, the probability of null-work is $P(w=0)=\sum_n\frac{e^{-\beta E_n}}{Z}|\langle E_n|\hat{U}(t)|E_n\rangle|^2$ with $|E_n\rangle$ being the $n$th energy level of $\hat{H}(\lambda)$ and $Z=\sum_ne^{-\beta E_n}$ the partition function. But in the Loschmidt echo perspective, the probability used to detect DQPT is $\mathcal{L}=|\sum_n\frac{e^{-\beta E_n/2}}{\sqrt{Z}}\langle E_n|\hat{U}(t)|E_n\rangle|^2=P(w=0)+\sum_{n,m\neq n}\frac{e^{-\beta (E_n+E_m)/2}}{Z}\langle E_n|\hat{U}(t)|E_n\rangle\langle E_m|\hat{U}^\dag(t)|E_m\rangle$. It can be seen that Loschmidt echo has additional interference terms than null-work probability. These additional interference terms, namely quantum coherence, can enhance quantum fluctuations, enabling them to overcome thermal fluctuations, thus reviving DQPT. This quantum coherence enhanced effect indicates that quantum coherence plays a crucial role in DQPT.

\begin{figure}
\begin{center}
\includegraphics[width=8cm]{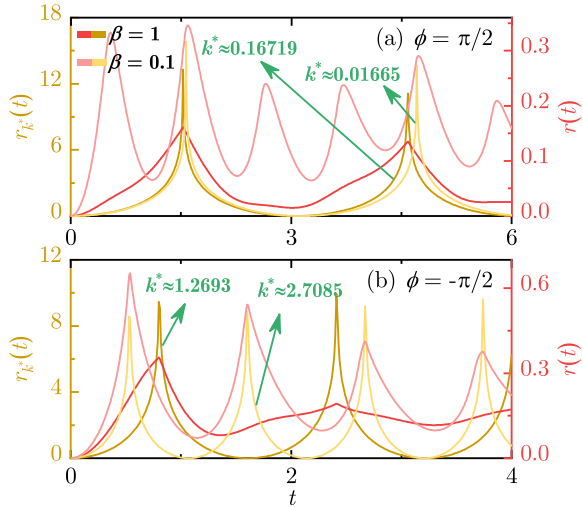}
\caption{(Color online) The time evolution of rate functions $r(t)$ (red) and the critical rate functions $r_{k^*}(t)$ (yellow) for the relative phases (a) $\phi=\pi/2$ and (b) $-\pi/2$ at temperatures $\beta=1$ (dark red and dark yellow curves) and $\beta=0.1$ (light red and light yellow curves). The critical modes used to obtain the critical rate function has been given (see the green arrows and values). The other parameters are $\lambda=0.5$, $\lambda'=2$ and $n=0$.}
\label{figure3}
\end{center}
\end{figure}

\section{The effects of quantum coherence on DQPT}
In this section, we investigate the effects of quantum coherence on DQPTs by letting $\phi=\pi/2$ and $-\pi/2$ as an example. At low temperature, the system mainly populates the ground state and has little quantum coherence, so this is not the case we want to discuss. At medial temperature $T\sim1$, the system will be excited to the higher energy levels, the quantum coherence between energy levels becomes significant in coherent Gibbs state. Under the influence of this quantum coherence, the line of Fisher zeros cuts $\mathrm{Im}(z)$ axis for the quench across the quantum critical point, and the critical mode depends on the relative phase $\phi$ [see the light blue and light red curves in Fig. \ref{figure2}(b1)]. The resulting nonanalytic behaviors of $r(t)$ and $\nu(t)$ can be observed which implies DQPT occurring [see the light blue and light red curves in Fig. \ref{figure2}(b2-b3)]. Otherwise, DQPT does not occur for the quench within the same phase [see the light blue and light red curves in Fig. \ref{figure2}(a1-a3) and (c1-c3)].

At high temperatures $T\sim10$, DQPT with the discontinuous jumps of $\nu(t)$ still occurs if the quench across the quantum critical point [see the dark blue and dark red curves in Fig. \ref{figure2}(b3)]. However, the rate function $r(t)$ does not always show cusp singularity. Specifically, the rate function for $\phi=-\pi/2$ shows cusp singularity, but it does not for $\phi=\pi/2$ [see the blue and red curves in Fig. \ref{figure2}(b2)]. To understand why some rate functions does not show cusp singularity, we still need to discuss the critical modes. We note that the critical modes for different relative phases and different quenches differ greatly, especially at high temperature. In the aforementioned discussion, we have pointed out that whether the critical mode dominates plays a crucial role on the behavior of $r(t)$. The rate functions for $\phi=\pm\pi/2$ are compared with their critical rate functions at different temperatures in Fig. \ref{figure3}. At medial temperature $\beta=1$, the rate functions for $\phi=\pi/2$ ($k^*\approx0.16719$) and $\phi=-\pi/2$ ($k^*\approx1.2693$) both resonate with their critical rate functions, thus the cusp singularity of rate function can be observed. At high temperature $\beta=0.1$, the critical mode for $\phi=\pi/2$ ($k^*\approx0.01665$) is drowned out, so that the cusp singularity of rate function can not be observed. But we can reverse the relative phase, i.e., $\phi=-\pi/2$, and get a critical mode ($k^*\approx2.7085$) close to $\pi$ which can survive at high temperatures. As a result, the cusp singularity of rate function for $\phi=-\pi/2$ can be observed at high temperature.

\begin{figure}
\begin{center}
\includegraphics[width=8cm]{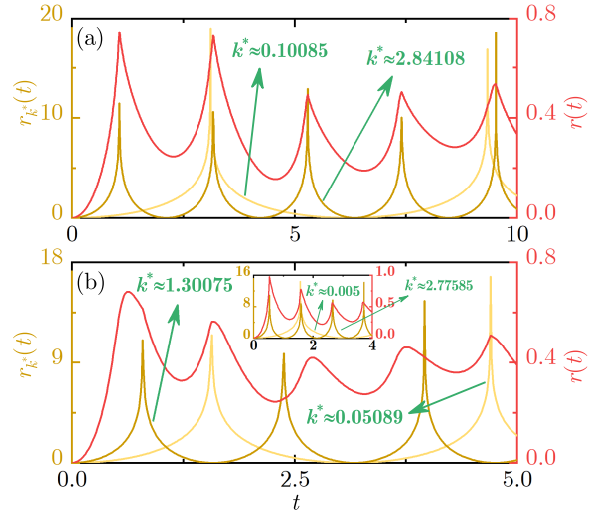}
\caption{(Color online) The time evolution of rate functions $r(t)$ (red) and the critical rate functions $r_{k^*}(t)$ (yellow) for a quench within (a) the ferromagnetic phase $\lambda=0\rightarrow\lambda'=0.5$ and (b) the paramagnetic phase $\lambda=1.5\rightarrow\lambda'=2$ at temperature $\beta=0.1$. The inset in (b) is the time evolution of rate functions $r(t)$ (red) and the critical rate functions $r_{k^*}(t)$ (yellow) for a quench within paramagnetic phase $\lambda=1.5\rightarrow\lambda'=2$ at temperature $\beta=0.01$. The critical modes used to obtain the critical rate function has been given (see the green arrows and values). The other parameters are $\phi=-\pi/2$ and $n=0$.}
\label{figure4}
\end{center}
\end{figure}

Very interestingly, some entirely new DQPTs independent of equilibrium quantum critical point can be observed. Specifically, when the transverse field is quenched in the ferromagnetic phase or paramagnetic phase at high temperature, the line of Fisher zeros cuts $\mathrm{Im}(z)$ axis twice, in other words there are two critical modes [see the dark blue curves in Fig. \ref{figure2}(a1) and (c1)], one approaches $\pi$ with temperature increasing, whereas another approaches $0$. The critical mode approaching to $\pi$ makes the winding number jump down but another approaching to $0$ makes it jump up [see the dark blue curves in Fig. \ref{figure2}(a3) and (c3)]. In Fig. \ref{figure4}, we plot these two critical rate functions, and compare them with rate function $r(t)$ at high temperature $\beta=0.1$. In the ferromagnetic phase, rate function $r(t)$ resonates with the critical rate function obtained by the critical mode closed to $\pi$, but detunes with another critical rate function. In other words, the critical mode closed to $\pi$ dominates at high temperatures, but the one closed to $0$ drowns out, thus only the cusp singularity in the rate function generated by the critical mode closed to $\pi$ can be observed. In the paramagnetic phase, both critical mode survive at $\beta=0.1$, thus two cusp singularities in the rate function can be observed. But if the temperature is increased to $\beta=0.01$, it is still the cusp singularity in the rate function generated by the critical mode closed to $\pi$ be observed [see the inset in Fig. \ref{figure4}(b)].

\section{Conclusions}
To summarize, the effects of quantum coherence on DQPTs have been studied by focusing the Fisher zeros, winding number and the rate function of Loschmidt echo after quenching the strength of transverse field in the one-dimensional quantum Ising model, where the quantum Ising model is initialized in the coherent Gibbs state. For the quench across equilibrium quantum critical point, DQPT will always occur at different temperatures, whether there is quantum coherence or not. As a result, the winding number will discontinuously jump down. Beyond that, we also found some entirely new DQPTs assisted by quantum coherence, which are independent of equilibrium quantum critical point. In these entirely new DQPTs, the line of Fisher zeros cuts $\mathrm{Im}(z)$ axis twice, i.e., there are two critical modes; one makes the winding number jump down but another makes it jump up. The rate function does not always show cusp singularity at high temperature because the critical mode no longer dominates. Quantum coherence assisted generation of DQPT implies that DQPTs originate from quantum fluctuations.

\section*{Acknowledgement}
This work was supported by the National Natural Science Foundation of China (Grant No. 11705099) and the Talent Introduction Project of Dezhou University of China (Grant Nos. 2020xjpy03 and 2019xgrc38).

\end{document}